\documentclass[useAMS,usenatbib]{mn2e}
\usepackage{graphicx}    
\usepackage{psfrag} 
\usepackage{natbib}
\usepackage{float}
\usepackage{amsmath}
\usepackage{caption}
\usepackage{subcaption}
\DeclareGraphicsExtensions{.ps,.pdf,.png,.eps}
\makeatletter
\def\fps@figure{htbp}
\makeatother

\newcommand*{\Msun}{\ensuremath{\mathrm{M_\odot}}}%

\begin{document}
\title[Impact of local density on galaxy sizes to $z\sim2$]{Evidence for a correlation between the sizes of quiescent galaxies and local environment to $z\sim2$}
\author[Lani et al.]{Caterina~Lani$^{1}$\thanks{E-mail:
ppxcl@nottingham.ac.uk}, Omar Almaini$^{1}$, William G. Hartley$^{1,2}$, Alice~Mortlock$^{1}$,
\newauthor  Boris H\"{a}u\ss{}ler$^{1,3,4}$, Robert W. Chuter$^{1}$, Chris Simpson$^{5}$, Arjen van der Wel$^{6}$,  
\newauthor Ruth Gr\"{u}tzbauch$^{1,7}$, Christopher~J.~Conselice$^{1}$, Emma J. Bradshaw$^{1}$, 
\newauthor  Michael C. Cooper$^{8}$, Sandra M. Faber$^{9}$, Norman A. Grogin$^{10}$, Dale D. Kocevski$^{11}$, 
\newauthor Anton M. Koekemoer$^{10}$ and Kamson Lai$^{9}$
\\
\footnotemark[0]\\
$^{1}$School of Physics and Astronomy, University of Nottingham, Nottingham, NG7 2RD, UK\\
$^{2}$Institute for Astronomy, ETH Zurich, Wolfgang-Pauli-Strasse 27, CH-8093 Zurich, Switzerland\\
$^{3}$University of Hertfordshire, Hatfield, Hertfordshire, AL10 9AB, UK \\
$^{4}$Department of Physics, University of Oxford, Denys Wilkinson Building, Keble Road, Oxford, Oxon, OX1 3RH, UK\\
$^{5}$Astrophysics Research Institute, Liverpool John Moores University, Twelve Quays House, Birkenhead, CH41 1LD, UK\\
$^{6}$Max-Planck Institut f\"ur Astronomie, K\"onigstuhl 17, D-69117, Heidelberg, Germany\\ 
$^{7}$Centre for Astronomy and Astrophysics, University of Lisbon, Portugal\\
$^{8}$Center for Galaxy Evolution, Department of Physics and Astronomy, University of California, Irvine, 4129 Frederick Reines Hall, \\ Irvine, CA 92697, USA\\
$^{9}$UCO/Lick Observatory, Department of Astronomy and Astrophysics, University of California, Santa Cruz, CA 95064, USA \\
$^{10}$Space Telescope Science Institute, 3700 San Martin Drive, Baltimore, MD 21218, USA \\
$^{11}$Department of Physics and Astronomy, University of Kentucky, Lexington, KY 40506, USA \\ }

\date{Accepted 2013 July 10. Received 2013 July 10; in original form 2013 March 8}
\pagerange{\pageref{firstpage}--\pageref{lastpage}} \pubyear{2013}
\maketitle

\label{firstpage}

\begin{abstract}
We present evidence for a strong relationship between galaxy size and
environment for the quiescent population in the redshift range $1<z<2$. Environments were measured
using projected galaxy overdensities on a scale of 400\,kpc, as determined
from $\sim96,000$ $K-$band selected galaxies from the UKIDSS Ultra Deep Survey
(UDS). Sizes were determined from ground-based $K-$band imaging,
calibrated using space-based CANDELS HST observations in the centre of the
 UDS field, with photometric redshifts and stellar masses derived from
11-band photometric fitting.  From the resulting size--mass relation, we
confirm that quiescent galaxies at a given stellar mass were typically $\sim50$\,\%
 smaller at $z\sim1.4$ compared to the present day.  At a given epoch, however, we
find that passive galaxies in denser environments are on average
significantly larger at a given stellar mass. The most massive quiescent
galaxies ($M_{*}>2\times$10$^{11}$\,M$_{\odot}$) at $z>1$ are typically 50\,\% larger in
the highest density environments compared to those in the lowest density environments.
Using Monte Carlo simulations, we reject the null hypothesis that the
size--mass relation is independent of environment at a significance $>4.8\,\sigma$ for the redshift range $1<z<2$. In contrast, the evidence for a relationship between size and environment is much weaker for star--forming galaxies.
\end{abstract}


\section{INTRODUCTION}
\label{sec:intro}
Numerous studies have tried to determine whether the evolution of galaxies and their properties are more heavily dictated by internal processes or environment, the so called ``nature versus nurture'' problem. Many galaxy properties (e.g. morphology, galaxy colour) appear to be related with environment but it has been difficult to disentangle the cause of these correlations, and whether they are produced by environmental processes. 
\newline \indent For decades it has been known that the morphology of galaxies in the local Universe is strongly related to environment. For example, \cite{Dressler:1980wq} studied 55 nearby rich clusters and found that the fraction of elliptical galaxies rises sharply with increasing density, while the corresponding fraction of spiral and irregular galaxies falls.  This differential spatial distribution for galaxies with different morphologies is known as the morphology--density relation \citep[e.g.][]{Oemler, Dressler:1980wq}.
\newline \indent  At low redshift, \cite{vdW2008} and \cite{Bamford2008} found that morphology, structure and colour are mainly dictated by galaxy stellar mass. If a fixed stellar mass is considered, however, they found that structure, morphology and colour all depend on environment. At higher redshift ($z\sim1$), using an optically selected sample, \cite{Cooper2006} found that the $(U-B)$ rest-frame colour is strongly dependent on environment: bluer galaxies generally live in less dense regions but their local mean density increases with luminosity. However, by going to a slightly higher redshift ($z\sim1.3$) and  using an optically selected sample, \cite{Cooper2007} suggested that blue and red galaxies inhabit indistinguishable environments. Conversely, a more recent study by \cite{Chuter2011} confirmed that galaxy colour is strongly related to the local density, at least out to $z\sim1.75$. In their work, which was based on a near-infrared selected sample, passive/red galaxies were found to inhabit denser environments than star--forming/blue galaxies. Moreover, the most luminous blue galaxies at $z\sim1$ appeared to live in environments which are as dense as the environments of red and passive systems at the same redshift. Several studies \citep[e.g.][]{Daddi2003, Quadri2007, Hartley2008, Hartley2010, Hartley2013}, based on deep near-infrared data, investigated larger scales through galaxy clustering. They found that red, passive galaxies are more strongly clustered than blue, star--forming galaxies out to at least $z\sim2$.
 \newline \indent Several studies were also undertaken considering star formation as a function of environment. In the local Universe, the environment was found to play an important role for star formation in galaxies, with the specific star formation rate decreasing sharply with local density (star formation--density relation, e.g. \citealt{Kauffmann2004}). In apparent contrast, at $z\sim1$ \cite{Sobral2011} found that the median star formation activity increases as a function of local surface density in group and field environments. However, once the highest densities are reached, the star--forming activity decreases strongly.
\newline \indent The primary motivation of our work is to understand whether galaxy environments are related to a particularly puzzling aspect of galaxy evolution, which is the apparent growth in galaxy size. Massive ($>10^{11}\rm\,M_{\odot}$) passive spheroids have been observed to be approximately 2--4 times more compact than galaxies of the same stellar mass at the present day \citep[e.g.][]{Daddi2005, Trujillo2007, Fer2008, vanDokkum2008, McLure2012, Poggianti2012}. It is then natural to wonder what happened to the population of extremely compact quiescent galaxies which were present at high redshift in large numbers, but seem to be much rarer at the present day \citep[e.g][]{Poggianti2012}. There are mainly two theories: these galaxies have ``puffed up'' either via internal processes, such as AGN feedback \citep[e.g.][]{Fan2008, Fan2010}, or via minor (in particular dry) mergers \citep[e.g.][]{Khochfar2006,Bournaud2007,Naab2009}. The latter has been supported by an increasing number of studies \citep[e.g.][]{Trujillo2011,Bluck2012,McLure2012}.  \cite{Hopkins2010} have applied several size growth models to a sample of spheroids in order to investigate how, at different epochs, these objects would move along the size--mass relation and how they would compare with observed galaxy properties. They concluded that later time major or minor dry merging, with lower density galaxies, is the dominant effect. They also conclude, however, that additional factors may be at work, such as equal density dry mergers, adiabatic expansion and also biases in the estimation of stellar masses.
\newline \indent Identifying a correlation between galaxy size and environment could help to explain the observed galaxy size evolution with redshift. This is because some of the key processes which affect the structure of galaxies take place in high densities. For example, the merger rate in intermediate densities is believed to be higher \citep[e.g.][]{Fakhouri2009,Lin2010,Lotz2011}. A number of studies have tried to investigate whether galaxy structural properties, such as size and morphology, depend on galaxy density. At low redshift, \cite{David2010} identified a weak dependence of the size--mass relation on local galaxy environments for low mass spiral galaxies.  In the field, they identified a population of low mass spiral galaxies  ($<10^{10}\rm\,M_{\odot}$)  with mean effective radii 15--20\,\% larger than the semi major axes of similar spirals in the cluster. This trend may suggest that extended disks do not survive in extreme cluster conditions.
\newline \indent For passive early-type galaxies in the redshift range $0.2<z<1.1$, \cite{Huertas2012} found no dependence of the size--mass relation on environments ranging from field to groups. Conversely, \cite{Cooper2012}, who focused on the high stellar mass end of red sequence galaxies at $0.4<z<1.2$, found early-type galaxies in the top 15\,\% of the density distribution to have effective radii 25\,\% larger than galaxies at the bottom 50\,\% of the density distribution. Finally in a forming cluster at $z\sim1.6$ \citep{Tanaka,Papovich2010}, a study by \cite{Papovich2012} found evidence for a lack of compact (circularised effective radius $\leq1$\,kpc) objects compared to the field at a similar epoch. 
 \newline \indent In this work we present a new study of correlations between galaxy size and environment at high redshift ($z>1$). We use data from the deepest $\sim1$\,deg$^{2}$ near-infrared survey to date (Almaini et al. in prep.) combined with sizes calibrated from HST CANDELS \citep{Grogin2011,Koekemoer2011}.
\newline \indent In Section 2 we present the data used in the analysis. Section 3 describes our methods to measure environments and structural parameters. Section 4 presents the results and Monte Carlo simulations to determine the significance of our findings. Section 5 provides a discussion, with a  summary and conclusions in Section 6. Additional tests on the robustness of our conclusions are presented in Appendix A and B. Throughout this work we adopted the following cosmology: $\Omega_{m}=0.3$, $\Omega_{\Lambda}=0.7$ and $H_{0}=70\rm\,km\rm\,s^{-1}\rm\,Mpc^{-1}$.

\section{THE DATA SETS AND SAMPLES SELECTION}
\subsection{UKIDSS UDS}
\label{sec:UDS}
This work is based on the Ultra Deep Survey (UDS; Almaini et al in prep.), which is the deepest component of the UKIRT (United Kingdom Infra-Red Telescope) Infra-Red Deep Sky Survey (UKIDSS; \citealt{Lawrence2007}). The 8th UKIDSS data release was used for this study. The UDS covers 0.77\,deg$^{2}$ and the current limiting magnitudes (AB), within an aperture of 2\,arcsec, are 24.9, 24.2, 24.6 (5$\,\sigma$) in $J$, $H$, $K$ respectively. The UDS also benefits from a large array of comparable multi-wavelength data. $U-$band data obtained with CFHT Megacam (Foucoud et al. in prep). These reach the limiting magnitude $U=26.75$ (AB, 2\,arcsec RMS). $B$, $V$, $R$, $i^{\prime}$ and $z^{\prime}$ --bands data obtained in the Subaru-XMM Deep Survey (SXDS; \citealt{Furusawa2008}). These achieve the following depths (AB, 5$\,\sigma$, within a 2\,arcsec aperture): $B=27.6$, $V=27.2$, $R=27.0$, $i^{\prime}=27.0$ and $z^{\prime}=26.0$. Near-infrared data from the Spitzer Legacy Program (SpUDS, PI: Dunlop) which reach limiting magnitudes (AB, 5$\,\sigma$) of 24.2 and 24.0 at $3.6\,\mu\rm m$ and $4.5\,\mu \rm m$ respectively.  All of these were fundamental for the compilation of adequate photometric redshifts, stellar masses and rest-frame magnitudes. Furthermore, existing $X-$ray and radio data (\citealt{Ueda2008} and \citealt{Simpson2006} respectively) were also used to remove obvious AGN. 
\newline \indent The galaxy catalogue employed in this work is $K-$band selected. A magnitude completeness cut of $K_\text{AB}=24.4$ was applied, leaving a final sample of $\sim96,000$ galaxies. This magnitude cut was found from simulations to produce a completeness of $\sim99$\,\%. It was determined by inserting fake galaxies into the image and re-running \textsc{SExtractor} to determine the faction of successfully re-extracted galaxies as a function of magnitude. For more details we refer the reader to \cite{Hartley2013}.

\subsection{Photometric Redshifts, Stellar Masses and Rest-Frame Magnitudes}
\label{sec:phot quantities}
\begin{figure}
  \begin{center}
 \includegraphics[scale=0.2]{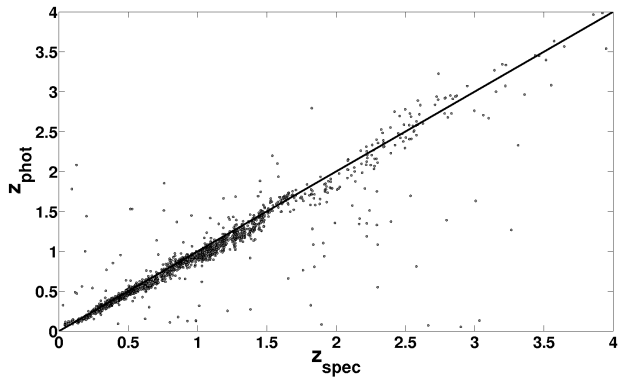}
\caption{Comparison between photometric and spectroscopic redshifts for approximately 2100 galaxies, with spectroscopic redshift obtained from a compilation of UDSz and archival data.}
\label{spec_phot}
\end{center}
\end{figure}

Photometric redshifts ($z_\text{phot}$) were determined by fitting template spectra to photometry from the following bands: $U$, $B$, $V$, $R$, $i^{\prime}$, $z^{\prime}$, $J$, $H$, $K$, $3.6\,\mu\rm m$ and $4.5\,\mu \rm m$. The package employed for the template fitting was \textsc{eazy} \citep*{EAZY2008}. The template fitting made use of the standard six \textsc{eazy} templates and an additional template, a combination of the bluest \textsc{eazy} template and a small amount of SMC-like extinction \citep{Prevot1984}. Furthermore, $\sim1500$ secure spectroscopic redshifts from the UDSz programme (an ESO Large Programme; PI: Almaini), and a few hundred archival spectroscopic redshifts were also used to train the fitting procedure (see \citealt{Simpson2012} and references therein for details of spectra used). A comparison of photometric and spectroscopic redshifts is shown in Figure \ref{spec_phot}. The dispersion between photometric and spectroscopic redshifts was measured to be $\delta z/(1+z)\sim0.031$. 
\newline \indent Stellar masses and rest-frame magnitudes were determined by employing a multi-colour stellar population fitting technique. This used a large grid of synthetic spectral energy distributions (SEDs) from the stellar population models of \cite*{Bruzual2003}, assuming a Chabrier IMF, to fit photometry from $U$, $B$, $V$, $R$, $i^{\prime}$, $z^{\prime}$, $J$, $H$, $K$, $3.6\,\mu\rm m$ and $4.5\,\mu\rm m$ --bands. The star formation history, with disparate ages, metallicity and extinctions, was modelled by an exponentially declining star formation, and parametrised by the onset of star formation and e-folding time as follows:

\begin{equation}
\label{SFH}
SFR(t)=SFR_{0}\times e^{-\frac{t}{\tau}}\,,
\end{equation}
where the e-folding time ranges between $\tau=0.01\,Gyr$ and $\tau=13.7\,Gyr$ and the age of the star formation onset ranges between $t=10^{-3}\,Gyr$  and $t=13.7\,Gyr$. Extinction due to galactic dust was modelled following \cite{Charlot2000}: the dust content was parametrised by $\tau_{v}$, the effective $V-$band optical depth, which was allowed to take values up to $\tau_{v}=5$. The fraction of extinction arising from dust in the 
inter-stellar medium was kept constant at 30\,\% (with the remaining extinction due to birth clouds which affects only stars with ages $< 
10^{7}\,yr$). The metallicity fraction was allowed to range between $Z=10^{-4}$ and $Z=0.1$. Templates were excluded if they were older than the age of the Universe at the redshift of the galaxy under consideration. The fitting procedure worked as follows. Firstly, all the synthetic SEDs in the grid were scaled, in the observed frame, to the $K-$band magnitude of the galaxy we wished to fit. Then each scaled template was fitted to the galaxy photometry resulting in a $\chi^{2}$ value. The best-fitting model template, together with the corresponding stellar mass and rest-frame magnitudes, were chosen according to the distributions of the resulting $\chi^{2}$ values. For more details on the production of photometric redshifts, stellar masses, stellar mass completeness and rest-frame colours we refer the reader to \cite{Hartley2013}.

\subsection{CANDELS/UDS}
\label{sec:CANDELS}
 The Cosmic Assembly Near-infrared Deep Extragalactic Legacy Survey (CANDELS; \citealt{Grogin2011}, \citealt{Koekemoer2011}) is
an on-going Hubble Space Telescope (HST) survey, carried out using Wide Field Camera 3 (WFC3) and the Advanced Camera for
Surveys (ACS). With 902 orbits it will cover a total area of
approximately 800 arcmin$^{2}$ and it consists of two sub-surveys: CANDELS Wide and
CANDELS Deep. CANDELS Wide consists of three fields, one of which is centered on the UDS (CANDELS--UDS). The imaging in $J$ (WFC3/IR filter F$_{125W}$) and $H$ (WFC3/IR filter F$_{160W}$) reach depths of $J=26.22$ and $H=26.32$ (AB, 5$\,\sigma$ and within 1\,arcsec$^{2}$; \citealt{Galametz2013}).  The relatively small area of CANDELS--UDS, however, did not offer sufficient dynamic range for our study of galaxy environments. We therefore measured environments using the larger ground-based UDS survey ($\sim$ 10 times the area of CANDELS--UDS), using the exquisite HST CANDELS imaging to calibrate the ground-based size measurements. Our method is outlined in \S \ref{sec:structural stuff}.

\begin{figure*}
\begin{center}
\includegraphics[width=180mm]{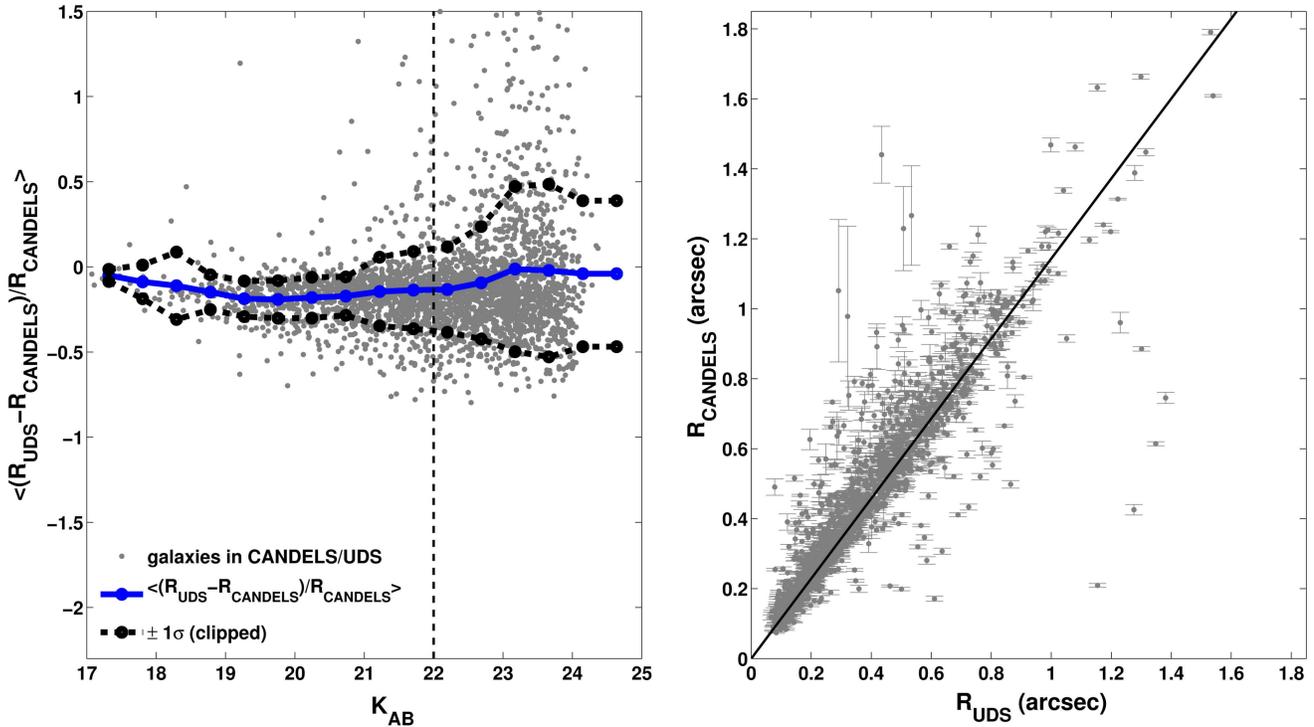}
\caption{{\textbf{Left panel:} fractional difference between effective radii measured from ground-based UDS $K-$band imaging and HST CANDELS $H-$band  imaging, as function of $K_\text{AB}$ magnitude. The mean values and standard deviations were measured on the clipped ($3\,\sigma$) distribution. Given the large dispersion and the much higher number of outliers for magnitudes fainter than $K_\text{AB}=22$, the effective radii were considered reliable for galaxies brighter than this magnitude. \textbf{Right panel:} comparison between effective radii measured from ground-based UKIRT data and CANDELS HST imaging, for galaxies with $K_\text{AB}\leq22$. The ground-based effective radii, in most cases, compare very well with the space-based effective radii though they show a systematic offset of 14\,\%. This offset is consistent with the findings of \protect \cite{Kelvin2012}, who compared galaxy size measurements in different wavebands.}}
\label{size_comparison}
\end{center}
\end{figure*}

\begin{figure*}
  \begin{center}
\includegraphics[width=180mm]{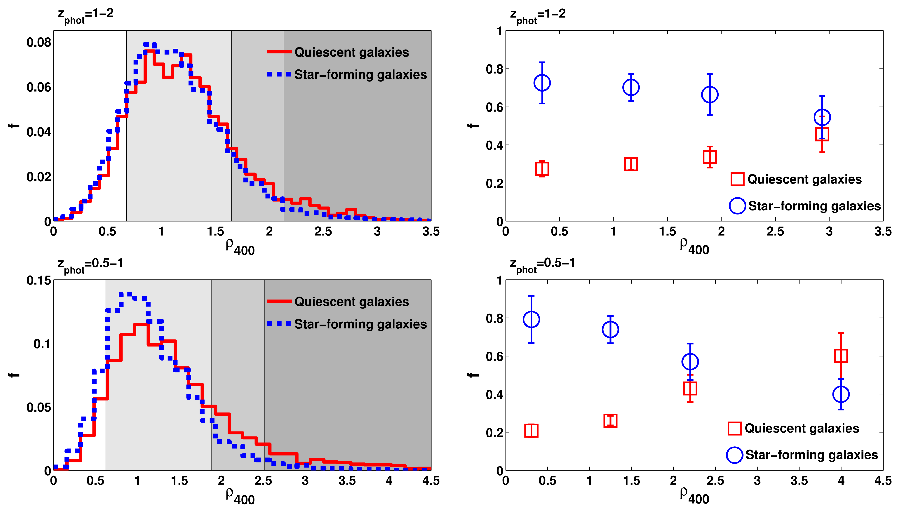}
\caption{{\textbf{Left panels:} histograms of galaxy density (as measured in a projected aperture of radius 400\,kpc) in the redshift ranges $1<z_\text{phot}<2$ (top) and $0.5<z_\text{phot}<1$ (bottom), separated into quiescent and star--forming galaxies. KS tests were performed on the density distributions. They rejected the null hypothesis that the quiescent and star-forming samples are drawn from the same underlying population at $\gg99.99\,\%$ confidence, for both $z_\text{phot}=1-2$ and $z_\text{phot}=0.5-1$. In these panels, shaded areas represent the four density bins considered in this work (see \S \ref{sec: colour density relation}). \textbf{Right panels:} fractions of quiescent and star--forming galaxies in the four density bins, on the top for $1<z_\text{phot}<2$ and on the bottom for $0.5<z_\text{phot}<1$. In both the redshift bins, the fraction of quiescent objects increases with density while the fraction of star-forming objects decreases with density.}}
\label{f_red_blue}
\end{center}
\end{figure*}

\section{METHOD}
In this section we present and discuss the key steps that were employed in our analysis. In \S \ref{sec:environments} we explain the techniques used to measure environments. In \S \ref{sec:structural stuff} we describe how the structural parameters were measured for the full UDS, and how these were calibrated against the structural parameters measured from CANDELS--UDS. In \S \ref{sec:quiescent vs SF} we describe the criteria employed to define samples of quiescent and star--forming galaxies, with particular emphasis in obtaining a strictly quiescent sample. 

\subsection{Environmental Measurements}
\label{sec:environments}
Environments were measured both using galaxy counts in a fixed physical aperture and distances to a range of $n^{th}$ nearest neighbours. 
For the former, a cylinder was constructed around the galaxy for which we wished to measure the local density. The radius of the cylinder was given by a fixed physical aperture size within which we wished to count galaxies; the depth of the cylinder was set to 1\,Gyr (in redshift space). This depth is several times the 1$\,\sigma$ error in the photometric redshifts and therefore minimises the exclusion of sources due to photometric redshift errors and, at the same time, avoids excessive dilution. The number count of real galaxies in an aperture, $N^\text{Aper}_\text{g}$, had then to be normalised in order to account for holes and edges in the field. This was done by measuring, $N^\text{Aper}_\text{Mask}$, the area of good pixels, i.e. pixels which were not masked due to the presence of a hole or the proximity to the field edge, within the chosen physical aperture. Moreover, $N^\text{Aper}_\text{g}$ also had to be normalised by the total number of galaxies over the field which lie within the considered 1\,Gyr redshift interval, $N_\text{z}$. The final density, $\rho_\text{aperture}$, for every galaxy in our catalogue was then calculated as follows:

\begin{equation}
\label{desity_ap}
\rho_\text{aperture}=\frac{N^\text{Aper}_\text{g}}{N_\text{z}}\times\frac{N^\text{Tot}_\text{Mask}}{N^\text{Aper}_\text{Mask}}\,,
\end{equation}
where $N^\text{Tot}_\text{Mask}$ is the total number of good pixels over the entire field.
\newline \indent In order to measure projected $n^{th}$ nearest neighbour densities it was necessary to first calculate projected distances, $d_\text{nth}$, to the $n^{th}$ nearest neighbour of interest. This was done by ranking projected distances, to all the galaxies in our sample, which lie within 1\,Gyr (in redshift space) centered on the galaxy for which the density was being measured. The final galaxy density was then calculated using the following equation:

\begin{equation}
\label{desity_nn}
\rho_\text{nth}=\frac{n}{\pi d^{2}_\text{nth}}\,,
\end{equation}
 where $n$ represents the $n^{th}$ nearest neighbour being considered. To account for holes and edges in the field, if the distance to the field edge or a masked region was less than the distance between the galaxy and the desired $n^{th}$ nearest neighbour then this object was discarded from the final analysis.
Of interest for this work is the study by \cite{Muldrew}, who applied several environment estimators to a common mock galaxy catalogue. Their findings show that the aperture method is a better probe of halo mass compared to $n^{th}$ nearest neighbour. For similar conclusions see also \cite{Haas2012}. In our work, we focused on environments measured by galaxy counts in an aperture, mainly for the following reason. The UDS field has a limited area, which presents actual holes and edges, rather than periodic boundaries. This made the use of the $n^{th}$ nearest neighbour technique very ineffective in terms of galaxy numbers, as many had to be discarded due to their position in proximity of either a masked region or the edge of the field, strongly weakening the statistics. Densities derived from $n^{th}$ nearest neighbour distances have, however, been used in this work as comparison (see \S\ref{sec: size density relation}).
\newline \indent Finally, it is important to stress that we cannot accurately measure the local environment of an individual galaxy with confidence, given the effects of photometric redshift dilution and projection effects. We aim, instead, to obtain robust statistical conclusions by comparing the average properties of large samples.

\subsection{Structural Parameters}
\label{sec:structural stuff}
Structural parameters were measured on the ground-based UDS $K-$band images using \textsc{galapagos} (Galaxy Analysis over Large Area: Parameter Assessment by \textsc{galfit}ing Objects from SExtractor; \citealt{Barden2012}). This makes use of both \textsc{SExtractor}, to identify and locate the objects to fit, and \textsc{galfit}, to fit S\`{e}rsic light profiles \citep{Sersic1968}. The shape of the S\`{e}rsic light profile is given in Equation \ref{Sersic_profile}
\begin{equation}
\label{Sersic_profile}
\Sigma(R)=\Sigma_\text{eff}\times \exp \left(-\kappa\left[\left(\frac{R}{R_\text{eff}}\right)^{1/n}-1\right] \right)\,,
\end{equation}
\noindent
 where $\Sigma$(R) is the surface brightness as a function of the radius R; $\Sigma_\text{eff}$ is the surface brightness at the effective radius, $R_\text{eff}$; $n$ is the S\`{e}rsic index; $\kappa$ is a function of $n$.
Since the UDS field is a mosaic, sixteen sub-regions (each corresponding to a single WFCAM camera chip) were fit separately, with a small overlap for the contiguous regions. In this procedure, the PSF was calculated locally, using $\sim100$ stars, within every sub-region. This approach, of considering each WFCAM camera chip separately, was used to tackle small PSF variations across the UDS ground-based mosaic.
\newline \indent The robustness of the ground-based sizes was addressed by comparing them to the CANDELS--UDS sizes, \citep{VanderWel2012} obtained from $H-$band data (Figure \ref{size_comparison}). This comparison showed that, although the space-based sizes are systematically 14\,\% larger, they are correlated with the ground-based sizes and were considered reliable for $K_\text{AB}\leq22$. Beyond this magnitude, both the number of outliers and the dispersion increase significantly. The 14\,\% offset is consistent with the offsets identified in \cite{Kelvin2012} when comparing size measurements obtained from different wavebands (see also \citealt{Hauessler2013}). 
\newline \indent When imposing a magnitude cut of $K_\text{AB}\leq22$, to ensure a highly complete  ($>95$\,\%) sample in stellar mass we required $\log\mathit{M_{*}/M_{\odot}}\geq9.8$ in the redshift range $z_\text{phot}=0.5-1$ and $\log\mathit{M_{*}}/M_{\odot}\geq10.45$ in the redshift range $z_\text{phot}=1-2$. Details on the mass completeness simulations can be found in \cite{Hartley2013}.
\newline \indent The fractional difference between ground-based and space-based sizes as a function of $K_\text{AB}$ magnitude is shown in the left panel of Figure \ref{size_comparison}. The right panel of Figure \ref{size_comparison}, compares ground-based and space-based sizes to our chosen limit of $K_\text{AB}=22$. To align the UDS sizes to the CANDELS sizes, the effective radii quoted from this point on were multiplied by a constant factor (i.e the gradient of the best-fit line in the right panel of Figure \ref{size_comparison} which was found to be 1.1431). This was done in order to allow an easier comparison with future studies based on CANDELS data.
\newline \indent Several tests were performed in order to check that the ground-based size measurements were robust in crowded regions, where inaccurate background subtraction could have occurred. These are described in Appendix A.

\subsection{Quiescent and Star Forming Populations}
\label{sec:quiescent vs SF}
For the purpose of this work we separated quiescent galaxies from star--forming (SF) galaxies using UVJ rest-frame colours, as described in \cite{StajinUVJ}. 
The criteria used in our study to select quiescent galaxies were taken from \cite{Williams2009}, who showed this  method was effective when applied to the first UDS data release. The required colours were found to be $(U-V)>1.3$, $(V-J)<1.6$ and in addition: \\
\begin{align*}
\label{UVJ}
(U-V)&> \left\{
\begin{array}{ll}
&0.88\times(V-J) + 0.59~~~~~~0.5<z<1.0  \,,\\
\\
&0.88\times(V-J) + 0.49~~~~~~1.0<z<2.0   \,.\\
\end{array}
\right.
\end{align*}
\\
To minimise the contamination from dusty SF galaxies which appear quiescent in $UVJ$ but are nevertheless forming stars, a maximum allowed specific star formation rate (sSFR, obtained from SED fitting; see \S \ref{sec:phot quantities}) was also considered. For the quiescent population, in addition to the UVJ selection, it was also required that galaxies have a $sSFR<7.43\times10^{-11}$\,yr$^{-1}$, that is to say a stellar mass doubling time longer than the age of the Universe. At the same time, the SF (i.e. non-quiescent) population is composed of all galaxies which did not make the quiescent category. 

\begin{figure*}
  \begin{center}
\includegraphics[width=180mm]{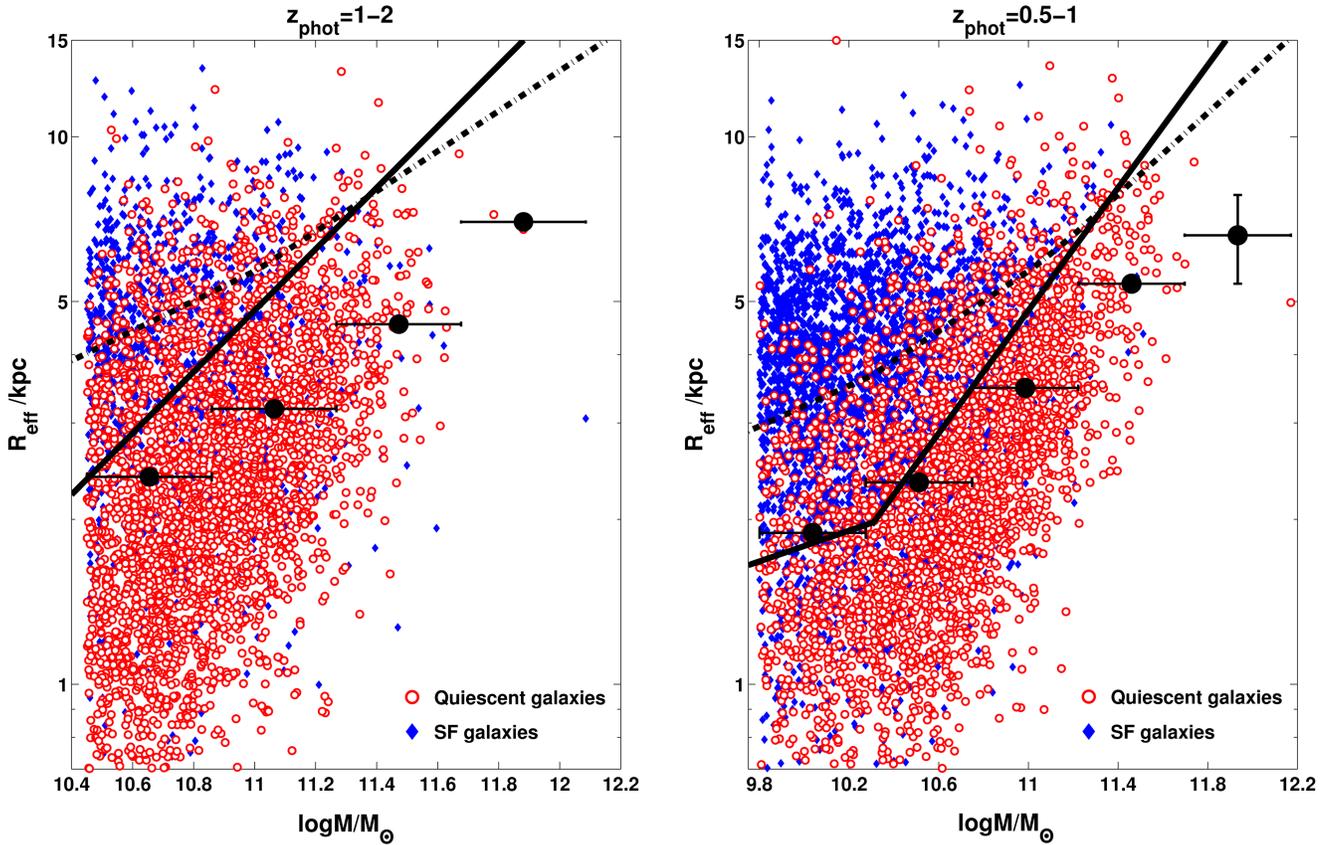}
\caption{\textbf{Left Panel:} effective radii versus stellar mass for quiescent and star-forming (SF) galaxies in the redshift range $1<z_\text{phot}<2$ ($\bar{z}~\sim1.4$). The black points correspond to the average effective radii for the quiescent population as measured in four stellar mass bins. The error bars in the $y-$direction are given by the standard error in the mean and in most cases are hidden by the data points themselves. We note that the black point corresponding to the most massive bin only contains two objects with similar effective radii. The black solid and dashed lines are the local relations for early-type galaxies (ETGs) and late-type galaxies (LTGs) respectively. These were modified from \protect\cite{Shen2003} to allow a comparison to our non-circularised effective radii (see \S \ref{sec: size mass relation}). Clear size evolution is apparent by comparison of our quiescent sample and the local ETG relation. Furthermore, at a given stellar mass quiescent galaxies are generally more compact than star-forming galaxies. \textbf{Right Panel:} effective radii versus stellar mass for quiescent and star-forming (SF) galaxies in the redshift range $0.5<z_\text{phot}<1$ ($\bar{z}~\sim0.7$). Here all the symbols have the same meanings as the symbols in the left panel, though in this case five stellar mass bins were considered in order to calculate the average effective radii for the quiescent galaxies. Massive quiescent galaxies are observed to be below the local ETG line and, once again, quiescent galaxies present smaller effective radii on average than star-forming galaxies at the same mass.}
\label{size_mass_both}
\end{center}
\end{figure*}

\section{Results}
\label{sec:results}
Throughout this section, unless otherwise stated, density is always measured in a projected 400\,kpc aperture radius. This aperture size was chosen as it compares well with the typical ``radius'' of clusters at high redshift. This aperture size was selected \textit{a priori}, but we find that our conclusions are not sensitive to the precise radius used (see \S\ref{sec: size density relation}).
 \subsection{The Colour--Density Relation}
\label{sec: colour density relation}
In this section we explore the colour--density relation and explain how our sample was binned, both in redshift and in density. The same binning will be used in this section, in \S\ref{sec: size mass relation} and in \S\ref{sec: size density relation}. 
\newline \indent Firstly our sample was split into two redshift bins, $1<z_\text{phot}<2$ and $0.5<z_\text{phot}<1$, both of which are approximately 2.6\,Gyr wide and include previously--identified clusters \citep{vanB,Geach2007,vanB2007,Papovich2010,Tanaka}. Four density bins, based on the full galaxy distribution in each redshift slice separately, were constructed as follows. Firstly, the mean and standard deviation of the density distribution (e.g. Figure \ref{f_red_blue}, top left panel) were calculated. Then bin 1($\equiv$``density1'') was constructed to contain galaxies with density more than $1\,\sigma$ below the mean of the density distribution. Bin 2 ($\equiv$``density2'') was made to include all the galaxies which have densities between $-1\,\sigma$ and $+1\,\sigma$ from the mean density, and therefore also contains the peak of the density distribution. Bin 3 ($\equiv$``density3'') was built to contain all objects with density between $+1\,\sigma$ and $+2\,\sigma$. The fourth density bin ($\equiv$``density4'') includes all the galaxies with density more than $2\,\sigma$ above the mean. These bins are over-plotted on the normalised density distributions in the top left panel (for $z_\text{phot}=1-2$) and bottom left panel (for $z_\text{phot}=0.5-1$) of Figure \ref{f_red_blue}. Here quiescent galaxies (red histograms) are shown to preferably inhabit denser environments than SF galaxies (blue histograms). As a result of KS tests on the normalised density distributions, the significance to which the SF and quiescent galaxies do not belong to the same underlying population is $\sim6.5\,\sigma$ (p--value $\sim1.5\times10^{-10}$) and $>10\,\sigma$ (p--value$\sim1.7\times10^{-62}$) for $1<z_\text{phot}<2$ and $0.5<z_\text{phot}<1$ respectively. A consistent pattern emerges in the right hand panels of Figure \ref{f_red_blue}, which show how the fraction of quiescent and SF galaxies vary in the four density bins described above. Here the fraction of quiescent galaxies increases as density increases, whereas the opposite trend is followed by SF galaxies (by construction). This is in very good agreement with previous results from \cite{Chuter2011}, who performed their study on the same field, but compared red and blue galaxies as defined by their rest-frame ($U-B)$ colour and absolute $K-$band magnitude information. 
\newline \indent It is worth noting that when producing the plots shown in Figure \ref{f_red_blue} no correction was applied to account for the colour--mass relation \citep[e.g.][]{Ruth2011} and the fact that the most massive galaxies are also the objects residing in the highest overdensities. Nonetheless, Figure \ref{f_red_blue} demonstrates that our measures of environment are sufficient to recover previously identified trends and separate the environments of passive and star--forming galaxies to $z\sim2$ with a high level of significance. 
\newline \indent Having obtained density measurements on which to build the rest of this work, we moved on to the study of galaxy sizes, mainly as a function of environment.

\begin{figure*}
  \begin{center}
\includegraphics[width=180mm]{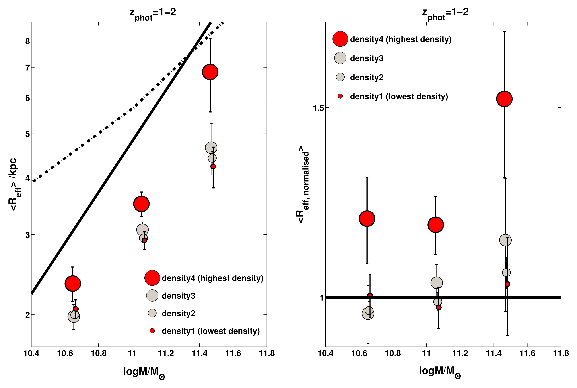}
\caption{{\textbf{Left Panel:} average sizes of quiescent galaxies as a function of stellar mass in the redshift range $1<z_\text{phot}<2$, separated into four density bins. Galaxy density was measured in fixed physical apertures of radius 400\,kpc. The black solid and dashed lines are the local relations for ETGs and LTGs respectively, taken from \protect\cite{Shen2003} and modified as described in \S \ref{sec: size mass relation}. Galaxies in the highest density bins appear significantly larger for the same stellar mass. \textbf{Right Panel:} mean normalised effective radii, obtained by first dividing out the size--mass relation, as described in \S\ref{sec: size density relation}. The fractional differences between the normalised effective radii of galaxies in the highest and lowest densities are (by increasing mass) $18\pm12$\,\%, $19\pm9$\,\% and $48\pm25$\,\%. For clarity, in both panels the data points were shifted by an arbitrarily small amount along the $x-$axis with respect to the centre of the stellar mass bins. For a list of the number of galaxies in each mass--density bin we refer the reader to Table \ref{tab:fig5}.}}
\label{Reff_mass_env12}
\end{center}
\end{figure*}

\begin{figure*}
  \begin{center}
\includegraphics[width=180mm]{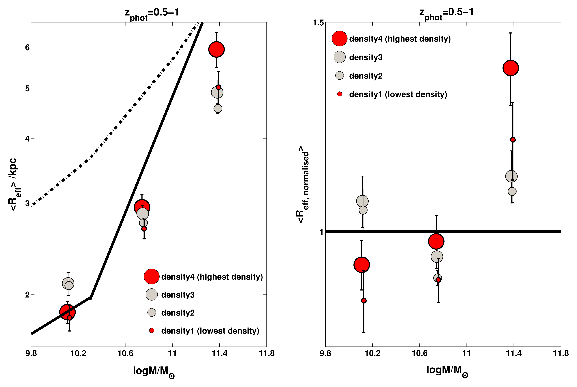}
\caption{{\textbf{Left Panel:} average sizes as a function of stellar mass in the redshift range $0.5<z_\text{phot}<1$, separated into four density bins. Galaxy density was measured in fixed physical apertures of radius 400\,kpc. The black solid and dashed lines are the local relations for ETGs and LTGs respectively, taken from \protect\cite{Shen2003} and modified as described in \S \ref{sec: size mass relation}. At high stellar masses, galaxies in the highest density bins appear larger than those in the lowest density bins, although the significance of the trends is weaker than those observed at $z_\text{phot}>1$ (Figure \ref{Reff_mass_env12}). \textbf{Right Panel:} mean normalised effective radii, obtained by first dividing out the size--mass relation, as described in \S\ref{sec: size density relation}. For clarity, in both panels the data points were shifted by an arbitrarily small amount along the $x-$axis with respect to the centre of the stellar mass bins. For a list of the number of galaxies in each mass--density bin we refer the reader to Table \ref{tab:fig6}.}}
\label{Reff_mass_env051}
\end{center}
\end{figure*}

\subsection{The Size--Mass Relation}
\label{sec: size mass relation}
Firstly we compared the size--mass relation of quiescent and SF galaxies in the redshift ranges $1<z_\text{phot}<$2 and $0.5<z_\text{phot}<$1 to the local relations \citep{Shen2003} for early-type and late-type galaxies (ETGs and LTGs respectively). When doing this we focused on the quiescent galaxies, $\sim2200$ objects in the redshift range $1<z_\text{phot}<$2 and $\sim2900$ objects in the redshift range $0.5<z_\text{phot}<$1. These broadly speaking, should be comparable to the local early-type galaxies in \cite{Shen2003}. Throughout this paper, however, the use of local relations for early-type and late-type galaxies was primarily to give a representation of the size--mass relation in the local Universe, rather than to constitute a direct comparison to our quiescent and SF populations. In both panels of Figure \ref{size_mass_both}, two salient features are immediately apparent. Firstly, in our sample we confirm the presence of size evolution for quiescent galaxies (\S\ref{sec:intro}), with a clear offset apparent between our redshift bins. Secondly, most of these objects lie below the local relations (black solid line for ETGs and black dashed line for LTGs). It is important to note that these local relations are a modified version of the local relations presented in \protect\cite{Shen2003}. A modification was necessary as, in \cite{Shen2003}, the sizes were measured within circular apertures whereas in this work the effective radii are not circularised. For a fairer comparison the local relations presented in \protect\cite{Shen2003} were therefore multiplied by the square root of typical axis ratios which, given the results from \cite{Padilla2008}, were set to be $\sqrt{0.75}$ for ETGs and $\sqrt{0.7}$ for LTGs. From Figure \ref{size_mass_both} it is also apparent that the quiescent population shows primarily smaller effective radii than the SF population at all redshifts considered in this work. Stronger size evolution for quiescent galaxies was also observed by other authors, such as \cite{Toft2007}, \cite{Fer2008}, \cite{Franx2008} and \cite{Williams2010}.
\newline \indent We then estimated the growth factor for the quiescent population, that is to say the typical growth needed to get onto the local ETG relation. For quiescent galaxies with stellar mass $>10^{11}\,\Msun$ and $z_\text{phot}=1-2$ ($\bar{z}~\sim1.4$), the growth factor was estimated to be $\sim100$\,\%. For quiescent galaxies with stellar mass $>10^{11}\,\Msun$ and $z_\text{phot}=0.5-1$ ($\bar{z}~\sim0.7$), the growth factor was estimated to be $\sim40$\,\%. Considering the wide range of sample selections and data used in the literature, our estimates are in broad agreement with previous works \citep[e.g.][]{Cimatti2008, VanderWel2008, McLure2012}. For quiescent galaxies with stellar mass $<10^{11}\,\Msun$, the growth factor was measured to be approximately 80\,\% for $1<z_\text{phot}<$2 ($\bar{z}~\sim1.4$) and 20\,\% for $0.5<z_\text{phot}<1$ ($\bar{z}~\sim0.7$), with generally a smaller growth factor corresponding to a lower stellar mass. It is worth noting that a growth of 80\,\% is in good agreement with the recent work by \cite{Poggianti2012}, although we note that other studies do not find the size growth to depend on stellar mass \citep[e.g.][]{Damjanov2011,Newman2012}. 
\newline \indent Overall, we broadly confirm the size growth for passive galaxies that has been observed in previous studies. A detailed determination of size evolution is beyond the scope of this paper and will be presented in van der Wel et al. (in preparation). The relationship between the local galaxy density and the sizes of quiescent and SF galaxies are described in the next section.

\begin{figure}
  \begin{center}
 \includegraphics[scale=1.2]{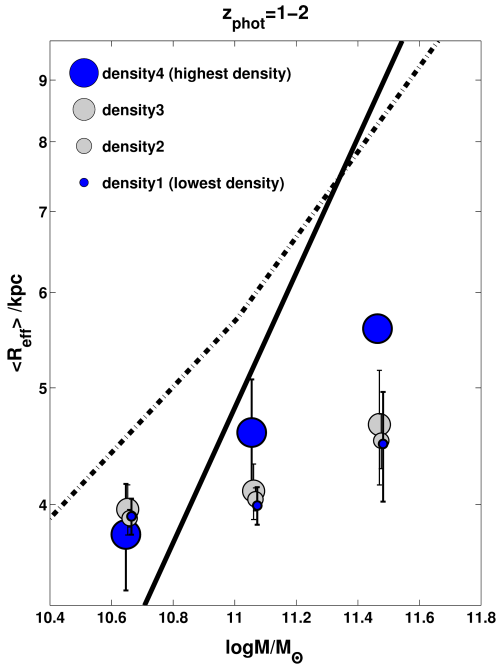}
\caption{{Mean sizes as a function of stellar mass for star-forming (i.e. non-passive) galaxies in the redshift range $1<z_\text{phot}<2$. Here all the symbols have the same meaning as Figure \ref{Reff_mass_env12}. The stellar mass and density bins are also analogous to the bins in Figure \ref{Reff_mass_env12}. The relation between size and density for the star--forming population is weaker than observed for the quiescent population. We highlight that, in the most massive bin, the ``density4'' point is only based on one galaxy. For a list of the number of galaxies in each mass--density bin we refer the reader to Table \ref{tab:fig7}.}}
\label{Reff_mass_env_blue_12}
\end{center}
\end{figure}

\begin{figure}
  \begin{center} 
\includegraphics[scale=1.32]{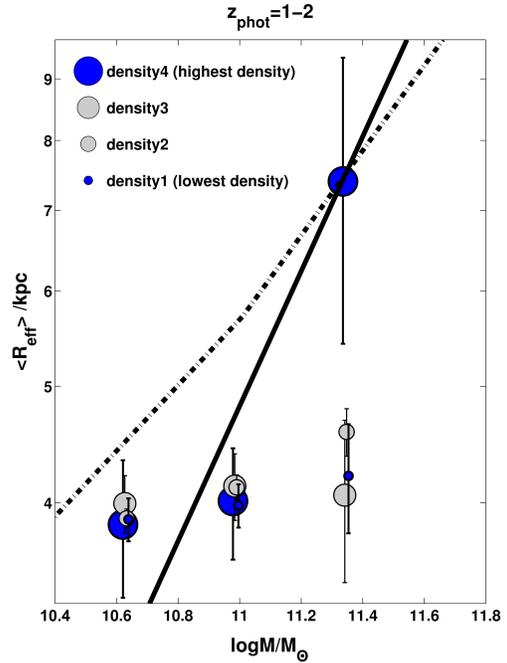}
\caption{{Mean sizes as a function of stellar mass for a more strictly defined sample of star-forming (see \S \ref{sec: size density relation}) galaxies in the redshift range $1<z_\text{phot}<2$ . Here all the symbols have the same meaning as Figure \ref{Reff_mass_env_blue_12}. There is no evidence for a relationship between size and environment of star--forming galaxies. Once again we highlight that, in the most massive bin, the ``density4'' point is only based on two galaxies. For a list of the number of galaxies in each mass--density bin we refer the reader to Table \ref{tab:fig8}.}}
\label{Reff_mass_env_strict_blue_12}
\end{center}
\end{figure}

\subsection{The Influence of Environment on the Size--Mass Relation}
\label{sec: size density relation}
In this section we investigate the relationship between galaxy size and local environment. We consider the same redshift intervals and density bins as described in \S \ref{sec: colour density relation}. However, within each density bin we further subdivide into three bins
of stellar mass.  In \S \ref{sec: size mass relation} we selected four
mass bins for the redshift interval $z_\text{phot}=1-2$ and five for
the redshift interval $z_\text{phot}=0.5-1$ in order to sample the
average size--mass relation. In this section, however, we select three
wider bins to allow the sample to be further subdivided by density.
For the redshift range $1<z_\text{phot}<2$ these are:
$10.45<\log\mathit{M_{*}}/M_{\odot}\leq10.86$,
$10.86<\log\mathit{M_{*}}/M_{\odot}<11.27$ and
$\log\mathit{M_{*}}/M_{\odot}\geq11.27$. For the redshift range
$0.5<z_\text{phot}<1$ these are:
$9.8<\log\mathit{M_{*}}/M_{\odot}\leq10.43$,
$10.43<\log\mathit{M_{*}}/M_{\odot}<11.06$ and
$\log\mathit{M_{*}}/M_{\odot}\geq11.06$.  The lower stellar mass
boundaries correspond to the stellar mass completeness limit
(\S \ref{sec:structural stuff}).  We note that the highest mass bins
are slightly wider (in log space) in order to include  two very
massive galaxies in both redshift intervals while maintaining a
reasonable number of objects per bin.  In order to check that this
binning was not affecting our results we repeated all of the analysis
below after excluding the two most massive systems. This did not
change the significance of any results.
\newline \indent The left panel of Figure \ref{Reff_mass_env12} shows the main result of this work. Here the average effective radii for quiescent galaxies in the four density bins (\S \ref{sec: colour density relation}) are plotted as a function of stellar mass for $z_\text{phot}=1-2$. It is clear that the average size of quiescent galaxies correlates with environment, with the most massive objects in the highest densities showing $48\pm25$\% larger mean normalised effective radii than galaxies in the lowest densities (at the same stellar mass). This trend was found to weaken with decreasing redshift (Figure \ref{Reff_mass_env051}). For $z_\text{phot}=0.5-1$, the size difference between galaxies inhabiting high and low densities is less significant.
\newline \indent A possible concern is that the environmental relations are affected by the trend of size with stellar mass within a given stellar mass bin. To remove any such effect, the following investigation was performed. The best-fit line to the average trend of size with mass of each population (e.g. Figure \ref{size_mass_both}, black points) was determined using the least square method. This was then employed to divide out the size--mass relation from the full, un-binned distribution and obtain the normalised effective radii, $R_\text{eff,~normalised}$, such that:
\begin{equation}
\label{Normalisation}
\log(R_\text{eff,~normalised})=\log(R_\text{eff})-(b\times {\log M_{*}}+a)\,,
\end{equation}
\noindent
where $a$ and $b$ are the intercept and the gradient of the best fit line respectively. The values of $a$ and $b$ are given in Table \ref{tab:values}.
This was performed for each population and redshift slice separately. The right panels of Figures \ref{Reff_mass_env12} and \ref{Reff_mass_env051} show that the relation between normalised galaxy sizes and environment is still present, especially at $z_\text{phot}>1$. 
\newline \indent For further checks and tests to asses whether our trends were driven by systematic effects we refer the reader to Appendix A and B. 
\newline \indent Our findings were also confirmed when looking at alternative measures of environment. We repeated the above analysis but this time using a 250\,kpc aperture radius. Galaxies which, according to this smaller aperture, lived in the densest regions showed up to $\sim71\pm30$\,\% larger mean normalised effective radii than galaxies, at comparable stellar masses, living in the lowest densities. Furthermore an alternative environmental measure, $n^{th}$ nearest neighbour distances (\S \ref{sec:environments}), was also explored. When we used a number of neighbours which translates into average distances comparable to 250--400\,kpc at $z_\text{phot}=1-2$, such as 15, consistent trends were recovered. Conversely, when smaller scales were explored, with distances to 3$^{rd}$ or even 8$^{th}$ nearest neighbour, the size--density relation appeared comparably strong only for the galaxies in the highest stellar mass bin.  \newline \indent Finally, we also repeated the above analysis but this time using a simple UVJ colour selection, with no additional \textit{sSFR} cut (see \S \ref{sec:quiescent vs SF}). The purpose of this exercise was to check that our chosen colour selection for passive galaxies did not affect the results described above. The alternative quiescent definition was not found to significantly change the results of our work, the overall trend for the passive population remained (albeit slightly weaker in the lowest stellar mass bin). 
\newline \indent Figure \ref{Reff_mass_env_blue_12} shows the average effective radii for SF (i.e. non-quiescent) galaxies in the four density bins (\S \ref{sec: colour density relation}), as a function of stellar mass for $z_\text{phot}=1-2$. After inspecting this figure, we noted there was arguably a hint of a relationship between size and environment for SF galaxies. However, due to the very strict criteria to select quiescent galaxies, our SF sample possibly contains some passive galaxies. The presence of these quiescent galaxies which did not satisfy our very strict cut may be driving this trend. To test this, we repeated Figure \ref{Reff_mass_env_blue_12} with a stricter SF sample, obtained by requiring a $sSFR>1\times10^{-10}$\,yr$^{-1}$. Figure \ref{Reff_mass_env_strict_blue_12} shows that, when a more strictly defined SF population is used, there is no evidence for a relationship between size and environment.
\newline \indent Due to the lack of a clear trend for star--forming galaxies, from this point onwards we will concentrate on the quiescent population. In the following section we describe the Monte Carlo simulations which were performed in order to investigate the validity of the results on the quiescent population.

\subsection{Monte Carlo Simulations}
\label{sec: MC}
To determine the significance of the difference between the sizes of quiescent galaxies in the highest and lowest densities, Monte Carlo (MC) simulations were performed as follows. Normalised effective radii were re-sampled, hundreds of thousands of times (with replacement), from galaxies in the lowest density bin for each of the three stellar mass ranges. These values were then used to calculate the probability of obtaining values equal to (or larger than) the mean sizes for galaxies in the highest density bins. The three probabilities were then multiplied together to calculate an overall probability for the re-sampled ``density1'' objects to be, on average, as large as ``density4'' objects. The overall probabilities are: $6.4\times10^{-10}$ ($\sim6.2\,\sigma$, assuming a normal probability distribution) and $9.6\times10^{-5}$ ($\sim4\,\sigma$, assuming a normal probability distribution) for galaxies in the redshift ranges $1<z_\text{phot}<2$ and $0.5<z_\text{phot}<1$ respectively. 
\newline \indent Since our MC simulations described above do not include any information 
from the intermediate density bins we also performed a second set of Monte Carlo simulations, where we compared two density bins obtained by combining ``density4''$+$``density3'' objects and ``density2''$+$``density1'' objects. In this case we obtained a significance of $\sim4.8\,\sigma$ for the redshift interval $1<z_\text{phot}<2$, and $\sim3.7\,\sigma$ for the redshift interval $0.5<z_\text{phot}<1$. 
\newline \indent From these simulations, and by inspection of Figures 5 and 6, we conclude that passive galaxies at a given stellar mass appear significantly larger in dense environments at $z>1$. At $z<1$ the dependence on environment appears more marginal, particularly at low mass.


\section{Discussion}
\label{sec:Discussion}
In this paper we investigate the relationship between environment and galaxy sizes to $z\sim2$. In this section we discuss our findings in the context of previous work, and how they fit into our wider understanding of galaxy formation and evolution. 
\newline \indent At $z\sim0.2$, \cite{David2010} did not find a significant dependence of size on environment, especially for the elliptical population, which should be loosely comparable to our quiescent population. We do not believe these results are inconsistent with our findings, for two reasons. Firstly, in our work, the size--density relation appears to be stronger for galaxies with stellar mass $>10^{11}\,\Msun$. These objects are, however, outside the stellar mass range considered by \cite{David2010}, which had an upper limit of $10^{11}\,\Msun$. Secondly, we observe the size--density relation to weaken from $z_\text{phot}=2$ to $z_\text{phot}=0.5$. This might imply that any correlation between environment and galaxy size becomes weaker as we approach the present day. In fact, a stronger size--density relation at higher redshift is also in line with theoretical work by \cite{Maulbetsch2007} who, with $N-$body simulations, found a dependence between the mass assembly history of dark matter haloes and environment. At $z>1$, the dark matter halo mass accretion rate is 4--5 times larger in denser environments. Conversely from $z<1$, the trend is reversed, with a mass accretion 4--5 times larger in low density environments.
\newline \indent At higher redshift, our work is also broadly consistent with the results from \cite{Cooper2012}, whose study focused on a spectroscopic sample of early-type galaxies with stellar mass between $10^{10}\,\Msun$ and $10^{11}\,\Msun$ and median redshift of $z\sim0.7$. Their environments ranging from field to groups. They found early-type galaxies on the red sequence ($U-B>1$) which live in groups to be 25\,\% larger than early-type galaxies, with comparable stellar masses and S\`{e}rsic indices, which live in the field. Despite the many differences (e.g. spectroscopic versus photometric redshifts and morphology selection) between the sample considered in \cite{Cooper2012} and the sample considered in our work, the results are qualitatively consistent. In contrast to our work, three recent studies found apparently different behaviour. \cite{Huertas2012} studied the size--mass relation as a function of environment for $\sim$ 700 group and field quiescent ETGs in the redshift range $z=0.2-1$. In their study they found no dependence of the size--mass  relation on environment. \cite{Raichoor2012} considered a sample of 76 ETGs at $z\sim1.3$, living in a range of environments. They found a hint that early-type galaxies in clusters are more compact than those in the field (the opposite of our findings), albeit with a confidence of $\sim90$\,\% according to a KS test. At similar redshifts, \cite{Rettura2011}  found no evidence for a difference in the size--mass relations for a sample of 45 cluster and field galaxies. The origin of the apparent discrepancy between these three studies and our work is unclear, but we note that our work is based on a  much larger sample of over 5000 passive galaxies. Our most significant signal also arises at higher redshift, which was not probed by the studies of \cite{Huertas2012}, \cite{Raichoor2012} or \cite{Rettura2011}. Furthermore, as previously mentioned in \S \ref{sec: colour density relation}, our sample includes denser environments than groups \citep[c.f.][]{Huertas2012}. \cite{Zirm2012} focused on a protocluster at $z\sim2$ and found a hint of quiescent cluster galaxies being larger than field quiescent galaxies at comparable redshift. 
\newline \indent  Our findings are also consistent with several recent theoretical studies. \cite{Shankar2011} used the latest Munich semi-analytic hierarchical galaxy formation models \citep{Guo2011} to investigate galaxy properties such as age and size. They found a relation between host halo mass and galaxy half-light radius, where more massive dark matter haloes host galaxies with larger half-light radii. This is driven by the interactions which satellite galaxies undergo when falling into larger dark matter haloes, such as stripping. They assume that, once the dark matter haloes of the satellites are being disrupted, so are their stellar components which then accrete onto the central galaxy. Another theoretical work by \cite*{Oogi2012} also identified a correlation between galaxy size and host halo mass. They used \cite{DeLucia2007} semi-analytic models and assumed that the most effective size growth mechanism is consecutive minor mergers. The more massive a halo, the more frequent the minor merging will be. This is in line with the conclusions of this work where galaxies in the highest densities, typically inhabiting the most massive dark matter haloes, show larger effective radii. 
\newline \indent As previously mentioned (\S \ref{sec: size mass relation}), several studies \citep[e.g.][]{Franx2008, Cimatti2012} suggest that size evolution with redshift is stronger for more massive galaxies ($>10^{11}\,\Msun$). Such behaviour would be consistent with high-density local environments playing a major role in size evolution, as indicated by our findings, since massive galaxies inhabit the densest environments on average.
\newline \indent It is important to consider, however, that our results are also consistent with an accelerated evolution for galaxies which inhabit the densest environments, i.e. those in the highest mass dark matter haloes may form and evolve at earlier times. Earlier evolution for galaxies in denser environments is already believed to be related to the observed colour--density relation \citep[e.g.][]{Chuter2011}. In such scenarios, the environment itself need not necessarily influence size evolution, and indeed a number of studies have suggested that merging alone cannot explain the observed size evolution of early-type galaxies \citep[e.g.][]{Damjanov2009,Nipoti2012}. Other growth scenarios may also be at work, such as adiabatic expansion due to mass loss, and could indirectly lead to a correlation of size with environment if they occur at earlier times within the most massive dark matter haloes. We also note that there is a possibility for the trends observed in our work to be driven by faster quenching in high density environments \citep[e.g.][]{Cassata2013}.
\newline \indent Finally, whatever the physical cause of the observed size evolution, we argue that the underlying correlation is likely to be between halo mass and galaxy size. On the scales probed in our analysis, halo mass is strongly related to the number of satellites \citep[e.g.][]{Skibba2009,Muldrew}. A full investigation of the effects of halo mass will require a careful decoupling of large-scale clustering and small-scale halo occupation \citep[e.g.][]{Hartley2013}. These effects will be investigated further in future work.

\section{Summary}
\label{sec: Summary}
Using a large $K-$band selected sample of galaxies we present evidence for a correlation between the size of quiescent galaxies and their environment in the redshift range $z_\text{phot}=0.5-2$. Environments were measured using projected galaxy overdensities and the distance to a range of $n^{th}$ nearest neighbours. Sizes were determined from ground-based $K-$band imaging, calibrated using space-based $H-$band CANDELS HST observations. Photometric redshifts and stellar masses were determined from 11-band photometric fitting. The main results obtained in this work are the following:
\begin{itemize}
\item The colour--density relation was observed to hold at least up to $z_\text{phot}\sim2$, with quiescent galaxies inhabiting denser environments than SF galaxies on average.
\\
\item Size evolution with redshift was confirmed for the quiescent population. From $z_\text{phot}\sim1.4$ to the present day, the most massive galaxies are, on average, found to double in size at a fixed stellar mass.
\\
\item We find that passive galaxies in denser environments (on a scale of
250--400\,kpc) are significantly larger at a given stellar mass in the
redshift range $1<z_\text{phot}<2$. The most massive quiescent galaxies ($M_{*}>2\times$10$^{11}$ M$_{\odot}$) at these epochs have effective radii that are $\sim 50$\,\% larger in the
highest density environments compared to those in the lowest density
environments. Monte Carlo simulations are used to test the significance of these
findings. They rejected the null hypothesis that galaxy sizes in the
densest environments are consistent with those at low density at a
significance of 6.2$\,\sigma$ in the redshift range $1<z_\text{phot}<2$, dropping
to 4.0$\,\sigma$ in the redshift range $0.5<z_\text{phot}<1$. Using a more conservative test, the significance of these differences dropped to 4.8$\,\sigma$ and 3.7$\,\sigma$ respectively (see \S \ref{sec: MC})
\\
\item The size--mass relation for star--forming galaxies shows no clear dependence on environment.
\end{itemize}

\section*{acknowledgements}
We are deeply grateful to the staff at UKIRT who were fundamental in making UKIDSS such a successful project. We would also like to thank the CANDELS team for their work, and the referee for their careful report which helped to improve our paper. CL wishes to thank Stuart Muldrew and Carlos Hoyos for many useful discussions.

\bibliographystyle{mn2e}
\bibliography{firts_year_cits.bib}

\bsp  

\appendix
\section{Tests on the robustness of the results } 
In this section we a present number of tests we performed to asses whether our findings were driven by systematic effects. Size measurements are very sensitive to the background estimation and in some regimes, such as crowded regions, an accurate background estimation can be more difficult to obtain. We also investigate the impact of size evolution within redshift bins, and the impact of using morphological criteria to select quiescent ``spheroids''.

\begin{figure}
 \begin{center}
\includegraphics[scale=2.13]{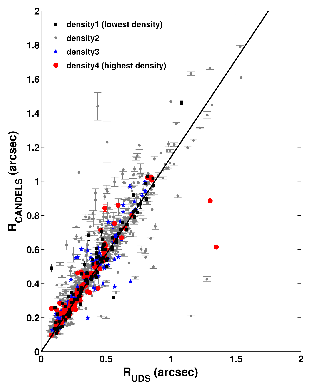}
\caption{Comparison between effective radii measured from ground-based UKIRT data and CANDELS HST data, for galaxies in the redshift range $0.5<z_\text{phot}<2$. The different symbols correspond to the  four density bins which were considered in our work. The effective radii measured on objects which inhabit the higher densities do not show a systematically less accurate ground-based size measurement.}
\label{size_comparison_density_coding}
\end{center}
\end{figure}

\subsection{Are size measurements behind dense structures reliable?}
The UDS field contains a particularly dense galaxy structure at $z\sim0.65$
\citep[e.g.][]{vanB}. We therefore performed tests to
determine whether the sizes of background galaxies were affected in
these crowded regions, due to potential difficulties in correctly estimating the sky background. Therefore the galaxies which lie behind (z$_{phot}>0.7$) the densest regions ($\rho_{400}\geq2.5$) of the aforementioned structure were excluded from our analysis. This excluded $\sim6$\,\% of quiescent galaxies in the redshift range $1<z_\text{phot}<2$. We then performed MC simulations on our smaller sample, focusing on quiescent galaxies with $z_\text{phot}=1-2$. The probability for the mean normalised half-light radii of galaxies in the lowest density regions to be larger than mean normalised half-light radii of galaxies in the highest density regions was then found to be $1.9\times10^{-10}$ ($\sim6.4\,\sigma$, assuming a normal probability distribution). This is formally of higher significance than the 6.2$\,\sigma$ signal obtained for the full sample (see \S \ref{sec: size density relation}). Furthermore, galaxies which lie behind the foreground structure
showed no evidence for a different distribution in galaxy
sizes. Comparing the half-light radii of these galaxies behind the
 dense structure with the rest of the field, a KS test did not
 reject the null hypothesis that the two subsamples belong to the
 same underlying population, and returned a p--value of $\sim0.5$.
These tests were also considered effective in checking that foreground dense structure was not enhancing the observed correlation of size with local density due to lensing.

\subsection{Are size measurements for galaxies with nearby neighbours reliable?}
Due to difficulties in subtracting the background correctly, objects with very close neighbours could potentially have less accurate size measurements. In order to tackle this problem in a simple and effective way, we repeated the analysis described in \S \ref{sec: size density relation}  after discarding galaxies with one or more neighbours within 2\,arcsec  (independent of redshift). Although this reduced the number of galaxies in our sample, particularly for galaxies in high density environments, the overall trend remained (albeit at lower significance). 
\newline \indent Furthermore, we repeated Figure \ref{size_comparison} (right panel) for galaxies in the redshift range $z_\text{phot}=0.5-2$. This time we used different symbols for galaxies inhabiting the four density bins considered throughout our work. This is shown in Figure \ref{size_comparison_density_coding}. Here objects which live in high-density environments do not exhibit systematically less accurate ground-based size measurements. More tests, also aiming to verify the robustness of size measurements in different densities, were performed by \cite{Boris2007} and \cite{Barden2012}. They did not find structural parameters to be less reliable in high densities.

\subsection{Does size evolution with redshift impact on our results?}
We also aimed to determine whether size evolution with redshift enhanced our results. This was a concern because the redshift bins considered throughout this work are not small and the redshift distributions for the galaxies in our sample, which inhabit different environments, are not identical. To account for the size evolution between $z_\text{phot}=2$ and $z_\text{phot}=1$ the following exercise was performed. Firstly the size increase with redshift for the quiescent population was estimated in each stellar mass bin separately. Mass bins were considered separately as size evolution may vary with stellar mass. The resulting fits were then used to normalise the sizes of quiescent galaxies at $z_\text{phot}>1$ to the value they would be if they were at $z_\text{phot}=1$. The effective sizes normalised to $z_\text{phot}=1$ were once again found to be, on average, larger for ``density4'' galaxies than for ``density1'' galaxies. Following MC simulations on the doubly normalised (for redshift and for stellar mass) radii, the probability for the mean half-light radii of quiescent galaxies in the lowest density regions to be equal or larger than the mean half-light radii of quiescent galaxies in the highest density regions was determined to be $3.43\times10^{-9}$ ($\sim5.9\,\sigma$, assuming a normal probability distribution), consistent with our original findings (\S \ref{sec: MC}). \newline \indent The same exercise was repeated for the quiescent population at $z_\text{phot}=0.5-1$. However this time the sizes were normalised to $z_\text{phot}=0.5$. In this case MC simulations returned a probability of $1.1\times10^{-3}$ ($\sim3.3\,\sigma$, assuming a normal probability distribution), which is also consistent with our original findings.
\begin{figure}
  \begin{center}
\includegraphics[scale=1]{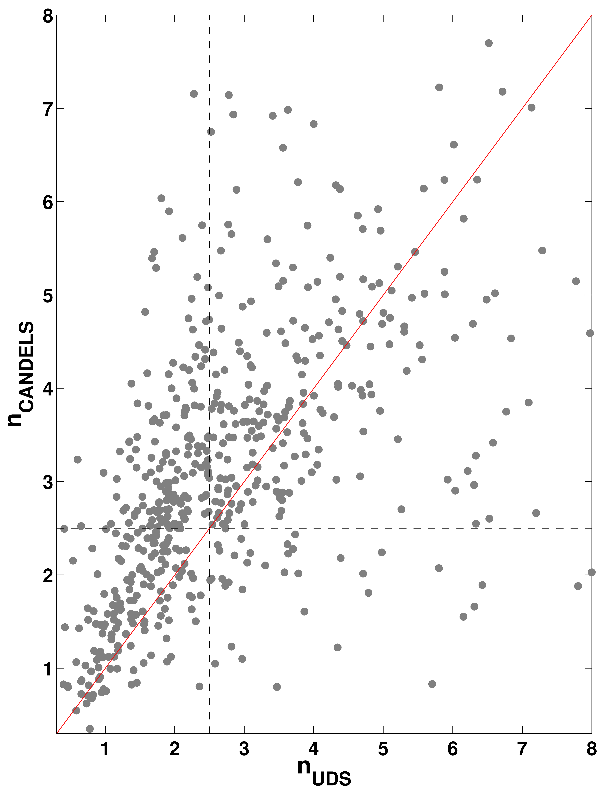}
\caption{{Comparison between S\`{e}rsic indices measured from ground-based UKIRT data ($K-$band) and CANDELS HST data ($H-$band), for quiescent galaxies with $K_\text{AB}<22$. In solid red is the 1:1 line, whereas the two black dashed lines show the boundary between ``disks'' and ``spheroids''. The contamination for the ``spheroid'' category, as defined according to the ground-based derived structural parameters, is much lower than the contamination for the ``disk'' category, as defined according to the ground-based derived structural parameters.}}
\label{n_comparison_quiescent}
\end{center}
\end{figure}

\begin{figure}
  \begin{center}
\includegraphics[scale=1]{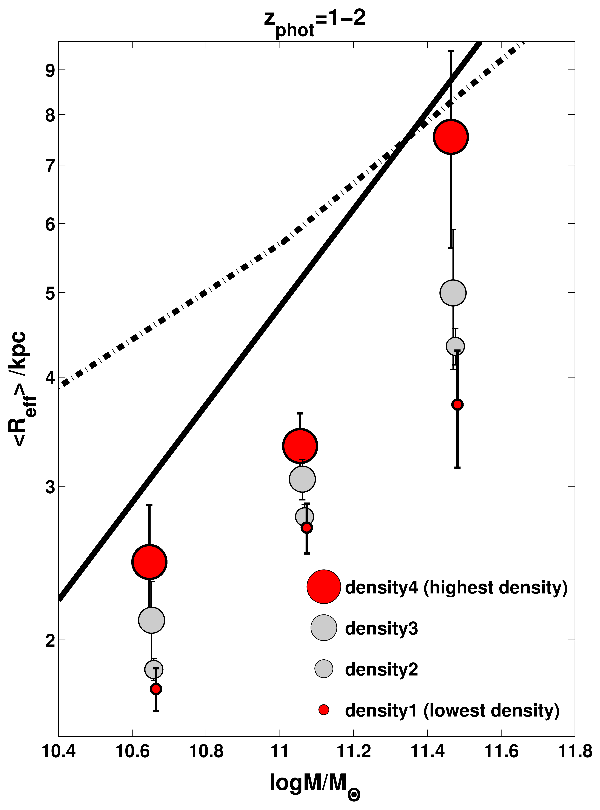}
\caption{{Mean sizes as a function of stellar mass in four density bins and in the redshift range $1<z_\text{phot}<2$, for quiescent ``spheroids'' only. Here all the symbols have the same meaning as in Figure \ref{Reff_mass_env12}. A relation between size and density is clearly present for the ``spheroidal'' population alone. For a list of the number of galaxies in each mass--density bin we refer the reader to Table \ref{tab:figA3}.}}
\label{Re_env_12_sph_only}
\end{center}
\end{figure}

\subsection{The influence of environment on the size--mass relation when only quiescent ``spheroids'' are considered} 
Recent studies \citep[e.g.][]{VdW2011} have shown that at $z\sim2$ a large fraction of quiescent galaxies are disk-dominated. Furthermore \cite{Basset2013} have found a hint of disk-dominated galaxies preferably inhabiting the in-fall region of the $z\sim1.6$ protocluster \citep{Papovich2010,Tanaka}. Therefore, if the effective radii of disk-dominated galaxies were biased towards larger values and if disk-dominated galaxies followed a morphology--density relation, our results could be affected. In order to tackle this problem we repeated Figure \ref{Reff_mass_env12} but only considering ``spheroids'' (S\`{e}rsic index $n\geq2.5$). Figure \ref{n_comparison_quiescent} compares S\`{e}rsic indices measured on space ($H-$band) and ground based ($K-$band) data for quiescent galaxies, with $K_\text{AB}<22$, in CANDELS--UDS. Despite the scatter in this relationship, the contamination
fraction for the ``spheroidal'' population, as defined according to the
ground-based derived structural parameters, is relatively small. Of
the galaxies classified as spheroids in the ground-based image, only 17\,\%
would be classified as ``disks'' using HST. On the contrary the contamination fraction for the ``disky'' population (S\`{e}rsic index $n<2.5$), as defined according to the ground-based derived structural parameters, is much larger ($\gg50$\,\%). For this reason we have only repeated Figure \ref{Reff_mass_env12} considering ``spheroids'' as we believe them to be more reliable than ``disks''. Figure \ref{Re_env_12_sph_only}  shows that a relationship between effective radii and local density of quiescent ``spheroids'' is obviously present. According to a MC simulation the probability that the mean half-light radii of quiescent ``spheroids'' in the lowest density regions are larger than mean half-light radii of quiescent ``spheroids'' in the highest density regions was determined to be $8.6\times10^{-12}$ ($\sim6.8\,\sigma$, assuming a normal probability distribution).

\section{Normalisation} 
In this section we test the normalisation of the average size--mass relation (as outlined in \S4.3), and report the best fit values employed through out our work.

\subsection{Robustness of our normalisation}
We investigated whether our normalisation (Equation \ref{Normalisation}) was robust both for our most massive and least massive galaxies. We repeated the same normalisation procedure (\S \ref{sec: size density relation}), but this time only considering the quiescent galaxies in highest and lowest stellar mass quartiles within each stellar mass bin. We found that the best fit lines to the average size--mass relation are virtually identical whether we use the full sample of passive galaxies or only the most massive and least massive passive galaxies within the three considered stellar mass bins.  
\newline \indent Furthermore, we also compared the mass distributions, for ``density1'' and ``density4'' quiescent galaxies, within the three considered mass bins (independently of environment). We found that the mass distributions are fully consistent as verified by KS tests.

\subsection{Best fit values employed in our work}
In table \ref{tab:values} we report the values obtained for the gradients and intercepts from Equation \ref{Normalisation} for our quiescent samples.

\begin{table*}
\renewcommand{\tabcolsep}{0.5cm}
\begin{tabular}{ c c c  }
 \hline\hline
    & $z_\text{phot}=1-2$      &     $z_\text{phot}=0.5-1$    \\
  \hline\hline  
      a   &   -4.39           &      -2.81    \\ [0.1cm]
      b   &   0.44          &        0.31   \\ 
\hline 
\end{tabular}
\caption{\textbf{Quiescent galaxies.} Values for gradients ($b$) and intercepts ($a$) of the best fit lines (Equation \ref{Normalisation}) to the average size--mass relation for the quiescent population in the two considered redshift bins.}
\label{tab:values}
\end{table*}

\section{Number of galaxies in the considered mass-density bins}
\begin{table*}
\caption{Number of galaxies considered in Figure \ref{Reff_mass_env12}.}
\renewcommand{\tabcolsep}{0.5cm}
\begin{tabular}{ c c c c }
\hline\hline
    & $10.45<\log\mathit{M_{*}}/M_{\odot}\leq10.86$      &     $10.86<\log\mathit{M_{*}}/M_{\odot}<11.27$     & $\log\mathit{M_{*}}/M_{\odot}\geq11.27$  \\
  \hline\hline  
      Density1 (lowest)   &   146        &      127       &       17    \\ [0.1cm]
      Density2   &   790       &        649     &       119   \\[0.1cm]
      Density3   & 122         &        122      &     18                      \\[0.1cm]
      Density4 (highest)   & 54          &        58        &      11             \\
\hline 
\end{tabular}
\label{tab:fig5}
\end{table*}

\begin{table*}
\caption{Number of galaxies considered in Figure \ref{Reff_mass_env051}.}
\renewcommand{\tabcolsep}{0.5cm}
\begin{tabular}{ c c c c }
\hline\hline
    & $9.8<\log\mathit{M_{*}}/M_{\odot}\leq10.43$      &     $10.43<\log\mathit{M_{*}}/M_{\odot}<11.06$     & $\log\mathit{M_{*}}/M_{\odot}\geq11.06$  \\
  \hline\hline  
      Density1 (lowest)   &   86        &      144       &       23    \\ [0.1cm]
      Density2   &   676       &        1170     &       293   \\[0.1cm]
      Density3   & 96         &        169      &     53                      \\[0.1cm]
      Density4 (highest)   & 80          &        93        &      29             \\
\hline 
\end{tabular}
\label{tab:fig6}
\end{table*}

\begin{table*}
\caption{Number of galaxies considered in Figure \ref{Reff_mass_env_blue_12}.}
\renewcommand{\tabcolsep}{0.5cm}
\begin{tabular}{ c c c c }
\hline\hline
    & $10.45<\log\mathit{M_{*}}/M_{\odot}\leq10.86$      &     $10.86<\log\mathit{M_{*}}/M_{\odot}<11.27$     & $\log\mathit{M_{*}}/M_{\odot}\geq11.27$  \\
  \hline\hline  
      Density1 (lowest)   &   206       &      126       &       12    \\[0.1cm] 
      Density2   &   1108       &        499     &       58   \\[0.1cm]
      Density3   &  128         &        71      &     10                      \\[0.1cm]
      Density4 (highest)   & 32         &        27        &      1             \\
\hline 
\end{tabular}
\label{tab:fig7}
\end{table*}

\begin{table*}
\caption{Number of galaxies considered in Figure \ref{Reff_mass_env_strict_blue_12}.}
\renewcommand{\tabcolsep}{0.5cm}
\begin{tabular}{ c c c c }
\hline\hline
    & $10.45<\log\mathit{M_{*}}/M_{\odot}\leq10.81$      &     $10.81<\log\mathit{M_{*}}/M_{\odot}<11.17$     & $\log\mathit{M_{*}}/M_{\odot}\geq11.17$  \\
  \hline\hline  
      Density1 (lowest)   &   156       &      103       &       16    \\ [0.1cm]
      Density2   &   840       &        444     &       75   \\[0.1cm]
      Density3   &  99        &        53      &     11                      \\[0.1cm]
      Density4 (highest)   & 23         &        20        &      2             \\
\hline 
\end{tabular}
\label{tab:fig8}
\end{table*}

\begin{table*}
\caption{Number of galaxies considered in Figure \ref{Re_env_12_sph_only}.}
\renewcommand{\tabcolsep}{0.5cm}
\begin{tabular}{ c c c c }
\hline\hline
    & $10.45<\log\mathit{M_{*}}/M_{\odot}\leq10.86$      &     $10.86<\log\mathit{M_{*}}/M_{\odot}<11.27$     & $\log\mathit{M_{*}}/M_{\odot}\geq11.27$  \\
  \hline\hline  
      Density1 (lowest)   &   72       &      65       &       10    \\ [0.1cm]
      Density2   &   368       &        376     &       75   \\[0.1cm]
      Density3   &  59         &        74      &     11                      \\[0.1cm]
      Density4 (highest)   & 19         &        32        &      7             \\
\hline 
\end{tabular}
\label{tab:figA3}
\end{table*}

In this section we list the number of galaxies in the considered mass--density bins. For clarity we provide a separate table for each one of the Figures showed in \S \ref{sec: size density relation} and Appendix A.

\label{lastpage}
\end{document}